\documentclass[aps,prd,onecolumn,groupedaddress,showpacs,nofootinbib,amssymb]{revtex4}
\usepackage[T1]{fontenc}
\usepackage[latin1]{inputenc}
\usepackage{graphicx}
\usepackage[english]{babel}
\usepackage{amsmath}
\usepackage{amssymb}
\usepackage{amsfonts}

\begin{document}

\def\pp{{\, \mid \hskip -1.5mm =}}
\def\cL{{\cal L}}
\def\be{\begin{equation}}
\def\ee{\end{equation}}
\def\bea{\begin{eqnarray}}
\def\eea{\end{eqnarray}}
\def\beq{\begin{eqnarray}}
\def\eeq{\end{eqnarray}}
\def\tr{{\rm tr}\, }
\def\nn{\nonumber \\}
\def\e{{\rm e}}

\title{Singularity of spherically-symmetric spacetime \\
in quintessence/phantom dark energy universe}

\author{Shin'ichi Nojiri$^1$ and Sergei D. Odintsov$^2$\footnote{
Also at Center of Theor. Physics, TSPU, Tomsk}}
\affiliation{
$^1$Department of Physics, Nagoya University, Nagoya 464-8602, Japan\\
$^2$Instituci\`{o} Catalana de Recerca i Estudis Avan\c{c}ats (ICREA)
and Institut de Ciencies de l'Espai (IEEC-CSIC),
Campus UAB, Facultat de Ciencies, Torre C5-Par-2a pl, E-08193 Bellaterra
(Barcelona), Spain
}

\begin{abstract}

We consider ideal fluid and equivalent scalar field dark energy universes
where all
four known types of finite-time, future singularities occur at some
parameter values. It is demonstrated that pressure/energy density of such
quintessence/phantom dark
energy diverges in spherically-symmetric spacetime at finite radius or at
the center. This may cause the instability of the relativistic star or
black hole in such universe. The resolution of the problem via the extra
modification of the equation of state is briefly discussed.

\end{abstract}


\maketitle

\section{Introduction}

The discovery of the late-time universe acceleration brought to the
playground the number of dark energy (DE) models with the effective
equation of
state (EoS) parameter $w$ being very close to $-1$ in accordance with
observational data. It is known that
 phantom/quintessence models lead to the violation of all/some of the energy
conditions. Such models unlike to $\Lambda$CDM with $w=-1$ lead to the
number
of quite surprising consequences in the remote future. For instance,
phantom DEs are characterized by the future Big Rip singularity\cite{Rip}.
Some of quintessence DEs bring the universe to softer finite-time
singularity in the future. Such finite-time future singularities may
represent so-called sudden singularities \cite{barrow, singularity} or
some other singularity types which are classified in
ref.\cite{Nojiri:2005sx}.
It is evident that the presence of finite-time future singularity in the
course of the universe evolution may show up at the current epoch.
One example has been given in ref.\cite{Kobayashi:2008tq} where it was
conjectured that sudden
singularity\cite{singularity} of specific modified gravity DE may make the
relativistic star formation process being unstable. The resolution of the
problem is to introduce the higher-order curvature terms
\cite{singularity} relevant only at the early universe in such a way that
future singularity disappears.

In the present work we consider the specific dark fluid which contains all
four known finite-time singularity types\cite{Nojiri:2005sx}. The
reformulation of it as scalar DE model with the same FRW asymptotic
solutions for the corresponding scalar potentials is also made.
The energy density/pressure of such singular DE may become divergent
in the spherically-symmetric spacetime at finite radius or at the center.
In a sense, that is the way the finite-time singularity
manifests itself as  singularity  of
spherically-symmetric space. This may lead to the instability of
relativistic
stars (in the same way as for modified gravity DE model in
ref.\cite{Kobayashi:2008tq}) or instability of black holes located in such
dark energy universe. It indicates that number of current DEs with such
properties may be problematic for realistic description of current
accelerating universe. Some extra EoS modification by the terms relevant
at the very early universe may be necessary in order to resolve this
problem.
Such modification is  discussed briefly in the last section.
The reconstruction method to find the specific dark energy
 responsible for any  singularity of spherically-symmetric space is also
presented.

\section{Finite-time singularities in the dark fluid universe with the
explicit equation of state}

In the present section we show the appearance of the finite-time
singularities in the
illustrative but sufficiently
realistic dark fluid with the following
 equation of state (EoS):
\be
\label{EOS1}
p = - \rho + A \rho^\alpha\ ,
\ee
with constant $A$ and $\alpha$.
We work in the spatially flat FRW space-time
\be
\label{fs0}
ds^2 = - dt^2 + a(t)^2 \sum_{i=1,2,3} \left(dx^i\right)^2\ .
\ee
Then since $H=\dot a/a$, the conservation law in the FRW universe,
$\dot \rho + 3 H \left( \rho + p \right) = 0$, gives
$\frac{d\rho}{3A \rho^\alpha} = - \frac{da}{a}$, which can be integrated as
\be
\label{EOS2c}
a = \left\{ \begin{array}{ll}
a_0 \e^{- \rho^{1-\alpha}/3A\left(1 - \alpha \right)}\ ,\quad & \alpha \neq 1 \\
a_0 \rho^{-1/3A}\ , \quad & \alpha = 1
\end{array} \right.
\ee
Here $a_0$ is a constant of integration.
In case $\alpha=1$, the EoS (\ref{EOS1}) reduces to the usual perfect fluid with
constant EoS parameter $w = -1 + A$. The universe is regular in the future
when $w\geq -1$ but there appears the Big Rip singularity when $w<-1$ for
the constant EoS parameter dark fluid.

The solution of the FRW equation
\be
\label{EOS2d}
\frac{3}{\kappa^2} H^2 = \rho\ ,
\ee
is given by
\be
\label{EOS2e}
a = \left\{ \begin{array}{ll}
\tilde a_0 t^{2/3A}\ ,\quad & \mbox{when}\ \alpha = 1\ ,\quad A>0 \\
\tilde a_0 \left( t_0 - t \right)^{-2/3A}\ ,\quad & \mbox{when}\ \alpha = 1\ ,\quad A<0
\end{array} \right. \ .
\ee
When $\alpha\neq 1$, by defining a new variable $N$, which is called  e-folding,
the FRW equation (\ref{EOS2d}) can be rewritten as
\be
\label{EOS2f}
N^{-1/2(1-\alpha)} \dot N = \left(\frac{\kappa^2}{3}\right)^{1/2} \left( - 3A
\left( 1 - \alpha \right)\right)^{1/2(1-\alpha)}\ .
\ee
Then in case $1/2(1-\alpha)=1$ or $\alpha = 1/2$, we find

\be
\label{EOS2h}
a = \tilde a_0 \e^{N_0\e^{- \frac{\sqrt{3}\kappa A t}{2}}}\ .
\ee

On the other hand, when $\alpha \neq 1/2$,
\be
\label{EOS2i}
\frac{2(1-\alpha)}{1 - 2\alpha} N^{\frac{1 - 2\alpha}{2(1-\alpha)}}
= \frac{\sqrt{3}\kappa A \left(t - t_0\right)}{2}\ .
\ee
Then the scale factor $a$ is given by
\be
\label{EOS2j}
a = \left\{ \begin{array}{l}
\tilde a_0 \e^{C t^\frac{2(1-\alpha)}{1-2\alpha}}\ ,\quad
C \equiv \frac{1}{1 - \alpha}\left(\frac{\kappa^2}{3}\right)^{\frac{1-\alpha}{1 - 2\alpha}}
\left( \frac{1}{2} - \alpha \right)^{\frac{2(1-\alpha)}{1 - 2\alpha}}
\left(- 3A\right)^{\frac{1}{1 - 2\alpha}} \\
\mbox{or}\ \tilde a_0 \e^{\tilde C \left(t_0 - t\right)^\frac{2(1-\alpha)}{1-2\alpha}}\ ,\quad
\tilde C \equiv \frac{1}{1 - \alpha}\left(\frac{\kappa^2}{3}\right)^{\frac{1-\alpha}{1 - 2\alpha}}
\left( \alpha - \frac{1}{2} \right)^{\frac{2(1-\alpha)}{1 - 2\alpha}}
\left(- 3A\right)^{\frac{1}{1 - 2\alpha}} \ .
\end{array} \right.
\ee
For the first case in (\ref{EOS2j}), one may put $t_0=0$ without loss of
the generality.
We may summarize the behavior of $a$ as follows,
\be
\label{EOS5}
a = \left\{ \begin{array}{ll}
\tilde a_0 t^{2/3A}\ ,\quad & \mbox{when}\ \alpha = 1\ ,\quad A>0 \\
\tilde a_0 \left( t_0 - t \right)^{-2/3A}\ ,\quad & \mbox{when}\ \alpha = 1\ ,\quad A<0 \\
\tilde a_0 \e^{N_0\e^{- \frac{\sqrt{3}\kappa A t}{2}}}\ ,\quad &
\mbox{when}\ \alpha = \frac{1}{2}\ ,\quad A<0 \\
\begin{array}{l}
\tilde a_0 \e^{C t^\frac{2(1-\alpha)}{1-2\alpha}}\ ,\quad
C \equiv \frac{1}{1 - \alpha}\left(\frac{\kappa^2}{3}\right)^{\frac{1-\alpha}{1 - 2\alpha}}
\left( \frac{1}{2} - \alpha \right)^{\frac{2(1-\alpha)}{1 - 2\alpha}}
\left(- 3A\right)^{\frac{1}{1 - 2\alpha}} \\
\mbox{or}\ \tilde a_0 \e^{\tilde C \left(t_0 - t\right)^\frac{2(1-\alpha)}{1-2\alpha}}\ ,\quad
\tilde C \equiv \frac{1}{1 - \alpha}\left(\frac{\kappa^2}{3}\right)^{\frac{1-\alpha}{1 - 2\alpha}}
\left( \alpha - \frac{1}{2} \right)^{\frac{2(1-\alpha)}{1 - 2\alpha}}
\left(- 3A\right)^{\frac{1}{1 - 2\alpha}}
\end{array} & \mbox{when}\ \alpha \neq 1,\ \frac{1}{2}
\end{array} \right.
\ee
Then the Hubble rate is given by
\be
\label{EOS6}
H = \left\{ \begin{array}{ll}
\frac{\frac{3}{2}A}{t}\ ,\quad & \mbox{when}\ \alpha = 1\ ,\quad A>0 \\
\frac{-\frac{3}{2}A}{t_0 - t}\ ,\quad & \mbox{when}\ \alpha = 1\ ,\quad A<0 \\
 - \frac{\sqrt{3}\kappa N_0 A }{2}\e^{- \frac{\sqrt{3}\kappa A t}{2}}\ ,\quad &
\mbox{when}\ \alpha = \frac{1}{2}\ ,\quad A<0 \\
\begin{array}{l}
\frac{2(1-\alpha)}{1 - 2\alpha}C t^{1/(1-2\alpha)} \\
\mbox{or}\ \frac{2(1-\alpha)}{1 - 2\alpha}\tilde C \left(t_0 - t\right)^{1/(1-2\alpha)}
\end{array} & \mbox{when}\ \alpha \neq 1,\ \frac{1}{2}
\end{array} \right.
\ee
Now one can describe the future, finite-time singularities of the universe
filled with above dark fluid.
When $\alpha<0$, there occurs Type II or sudden future singularity
\cite{barrow, singularity}.
When $0<\alpha<1/2$ and $1/(1-2\alpha)$ is not an integer, there occurs
Type IV singularity.
When $\alpha=0$, there is no any singularity.
When $1/2<\alpha<1$ or $\alpha=1$ and $A<0$, there appears Type I or Big Rip type
singularity.
When $\alpha>1$, there occurs Type III singularity.

The above general classification of singularities was first given in \cite{Nojiri:2005sx}:
\begin{itemize}
\item Type I (``Big Rip'') : For $t \to t_s$, $a \to \infty$,
$\rho \to \infty$ and $|p| \to \infty$. This also includes the case of
$\rho$, $p$ being finite at $t_s$.
\item Type II (``sudden'') : For $t \to t_s$, $a \to a_s$,
$\rho \to \rho_s$ and $|p| \to \infty$
\item Type III : For $t \to t_s$, $a \to a_s$,
$\rho \to \infty$ and $|p| \to \infty$
\item Type IV : For $t \to t_s$, $a \to a_s$,
$\rho \to 0$, $|p| \to 0$ and higher derivatives of $H$ diverge.
This also includes the case in which $p$ ($\rho$) or both of $p$ and $\rho$
tend to some finite values, while higher derivatives of $H$ diverge.
\end{itemize}
Here, $t_s$, $a_s (\neq 0)$ and $\rho_s$ are constants.

In case of Type II singularity, where $\alpha<0$, $H$ vanishes as
$H \sim \left(t_0 - t\right)^{1/(1-2\alpha)}$ when $t\to t_0$
and therefore we find $\rho$ vanishes as it follows from
the FRW equation (\ref{EOS2d}). Then, near the singularity, the EoS
(\ref{EOS1}) is reduced to
\be
\label{EoSt1}
p \sim A\rho^{\alpha}\ .
\ee
On the other hand, in case of Type I singularity, where $1/2<\alpha<1$
or $\alpha=1$ and $A<0$, $H$ and therefore $\rho$ diverge when $t\to t_0$. Then the EoS (\ref{EOS1})
reduces to
\be
\label{EoSt2}
p \sim - \rho\ \mbox{or}\ p\sim -(1-A)\rho\ .
\ee
In case of Type III singularity, where $\alpha>1$,
$H$ and  $\rho$ diverge when $t\to t_0$ and therefore the EoS
(\ref{EOS1})
reduces to
\be
\label{EoSt3}
p \sim A \rho^\alpha\ .
\ee
Hence, we illustrated the occurrence of all four types of future
singularity for the above dark fluid. Actually, the corresponding
singular asymptotics of such as well as of the more complicated EoS dark
fluid are
given in
ref.\cite{Nojiri:2005sx}.

\section{Finite-time singularities in the scalar field dark energy
universe}

We now consider the scalar-tensor theory, which also leads to the above
singularities.
One starts from the following action for the scalar-tensor theory;
\be
\label{EOS7}
S=\int d^4 x \left\{ \frac{R}{2\kappa^2}
 - \frac{\omega(\phi)}{2} \partial_\mu \phi \partial^\mu \phi - V(\phi)\right\}\ .
\ee
Here $\omega(\phi)$ and $V(\phi)$ are some functions of the scalar field
$\phi$.
Note that scalar field may be always redefined so
that kinetic function is absorbed.
In fact, in case $\omega(\phi)$ is positive definite, if we redefine the
scalar field $\phi$ by $\varphi = \int d\phi \sqrt{\left|\omega(\phi)\right|}$, we obtain
\be
\label{EOS7c}
S=\int d^4 x \left\{ \frac{R}{2\kappa^2}
\mp \frac{1}{2} \partial_\mu \varphi \partial^\mu \varphi - \tilde V(\varphi)\right\}\ .
\ee
Here the potential $\tilde V(\varphi)$ is defined by $\tilde V(\varphi)\equiv
V\left(\phi\left(\varphi\right)\right)$. The minus (plus) sign in $\mp$
corresponds
to positive (negative) $\omega(\phi)$.

Let us consider the theory in which $V(\phi)$ and $\omega(\phi)$ are given
by
\be
\omega (\phi ) = -\frac{2}{\kappa ^{2}}f^{\prime }(\phi )\ ,\qquad
V(\phi ) = \frac{1}{\kappa ^{2}} \left[ {3f(\phi)}^{2}+f^{\prime }(\phi ) \right]\ ,
\label{eq:1.6a}
\ee
where $f(\phi)$ is a proper function of $\phi$. Then the following solution
is found \cite{Nojiri:2005pu}
\be
\label{ST1}
\phi =t\ , \quad H(t)=f(t)\ .
\ee

For the action (\ref{EOS7}), the Hubble rate $H$ in (\ref{EOS6}) could be realized
by the following scalar models:
\bea
\label{EOS8a}
& \omega(\phi) = \frac{3A}{2\kappa^2 \phi_\pm^2} \ ,\quad
V(\phi) = \frac{\frac{27A^2}{4} - \frac{3A}{2}}{\phi^2} \ ,\quad &
\mbox{when}\ \alpha = 1 \\
\label{EOS8b}
& \omega(\phi) = - \frac{3}{2}N_0 A^2 \e^{-\frac{\sqrt{3}\kappa A \phi}{2}}\ ,\quad
V(\phi) = \frac{3A^2}{4}\left( 3 N_0^2 \e^{-\sqrt{3}\kappa A \phi}
+ N_0 \e^{-\frac{\sqrt{3}\kappa A \phi}{2}} \right)\ ,\quad &
\mbox{when}\ \alpha = \frac{1}{2}\ ,\quad A<0 \\
\label{EOS8c}
& \omega(\phi) = - \frac{4(1-\alpha) C \phi_\pm^{\frac{1}{1-2\alpha}}}{1-2\alpha}\ ,\quad
V(\phi) = \frac{2(1-\alpha)}{(1 - 2\alpha)^2 \kappa^2} \left\{
6(1-\alpha)C^2 \phi_\pm ^{\frac{2}{1-2\alpha}} + C \phi_\pm^{\frac{2\alpha}{1-2\alpha}}\right\} \ ,
& \mbox{when}\ \alpha \neq 1,\ \frac{1}{2}
\eea
Here $\phi_+ = \phi$ and $\phi_- = t_0 - \phi$. In (\ref{EOS8a}), $\phi_+$
($\phi_-$)
when $A>0$ ($A<0$).

Let us investigate the form of $\tilde V(\varphi)$ in (\ref{EOS7c})
when we rewrite the action (\ref{EOS7}).
\bea
\label{EOSAAa}
& \varphi = \sqrt{\pm \frac{3A}{2\kappa^2}}\ln\frac{\phi_\pm}{\phi_0}\ ,\quad
\tilde V(\varphi) = \left(\frac{27A^2}{4} - \frac{3A}{2}\right)
\frac{\e^{-2\kappa\varphi\sqrt{\pm\frac{2}{3A}}}}{\phi_0^2}\
& \mbox{when $\alpha = 1$} \\
\label{EOSAAb}
& \varphi= - \frac{\sqrt{\mp 8N_0}}{\kappa}\e^{- \frac{\sqrt{3}\kappa A}{4}\phi}\ ,\quad
\tilde V(\varphi) = \frac{3A^2}{4}\left(\frac{3\kappa^4}{64}\varphi^4
\mp \frac{\kappa^2}{8}\varphi\right)\
& \mbox{when $\alpha = \frac{1}{2}$} \\
\label{EOSAAc}
& \varphi = \sqrt{\pm 2C}\ln \frac{\phi}{\phi_0}\ ,\quad
\tilde V(\varphi) = \frac{2}{\kappa^2}\left\{\frac{3C^2}{2\phi_0^4}\e^{- \frac{4\varphi}{\sqrt{\pm 2C}}}
+ \frac{C}{\phi_0^3} \e^{- \frac{3\varphi}{\sqrt{\pm 2C}}} \right\}\
& \mbox{when $\alpha = \frac{3}{4}$} \\
\label{EOSAAd}
& \varphi = \mp \sqrt{\pm 2\tilde C}\ln \frac{t_0 - \phi}{\phi_0}\ ,\quad
\tilde V(\varphi) = \frac{2}{\kappa^2}\left\{\frac{3\tilde C^2}{2\phi_0^4}
\e^{ \frac{4\varphi}{\sqrt{\pm 2\tilde C}}}
+ \frac{\tilde C}{\phi_0^3} \e^{\frac{3\varphi}{\sqrt{\pm 2\tilde C}}} \right\}\
& \mbox{when $\alpha = \frac{3}{4}$}
\eea
Here $\phi_+ = \phi$ and $\phi_- = t_0 - \phi$, again, $\phi_0$ is the
constant of the integration,
the upper (lower) sign corresponds to $-$ ($+$) sign in (\ref{EOS7c}).
In (\ref{EOSAAa}), (\ref{EOSAAb}), (\ref{EOSAAc}), and (\ref{EOSAAd}),
the upper (lower) sign corresponds to $A>0$ ($A<0$), $N_0<0$ ($N_0>0$), $C>0$ ($C<0$),
$\tilde C>0$ ($\tilde C<0$), respectively.

We also find when $\alpha\neq 1,\frac{1}{2},\frac{3}{4}$,
\bea
\label{EOSa9a}
&& \varphi = \frac{4\sqrt{\mp \left(1-2\alpha\right)\left(1 - \alpha\right) C}}{3-4\alpha}
\phi^{\frac{3-4\alpha}{2-4\alpha}}\ ,\nn
&& \tilde V(\varphi) = \frac{2\left(1-\alpha\right)}{\left(1-2\alpha\right)^2\kappa^2}
\left\{ 6 \left(1-\alpha\right)C^2 \left(\frac{\left(3-4\alpha\right)\varphi}
{4\sqrt{\mp \left(1-2\alpha\right)\left(1 - \alpha\right) C}}\right)^{\frac{4}{3-4\alpha}}
+ C \left(\frac{\left(3-4\alpha\right)\varphi}
{4\sqrt{\mp \left(1-2\alpha\right)\left(1 - \alpha\right) C}}\right)^{\frac{4\alpha }{3-4\alpha}}
\right\}\ , \\
\label{EOSa9b}
&& \varphi = - \frac{4\sqrt{\mp \left(1-2\alpha\right)\left(1 - \alpha\right) \tilde C}}{3-4\alpha}
\left(t_0 - \phi\right)^{\frac{3-4\alpha}{2-4\alpha}}\ ,\nn
&& \tilde V(\varphi) = \frac{2\left(1-\alpha\right)}{\left(1-2\alpha\right)^2\kappa^2}
\left\{ 6 \left(1-\alpha\right)\tilde C^2 \left(- \frac{\left(3-4\alpha\right)\varphi}
{4\sqrt{\mp \left(1-2\alpha\right)\left(1 - \alpha\right) \tilde C}}\right)^{\frac{4}{3-4\alpha}}
+ \tilde C \left(- \frac{\left(3-4\alpha\right)\varphi}
{4\sqrt{\mp \left(1-2\alpha\right)\left(1 - \alpha\right) C}}\right)^{\frac{4\alpha }{3-4\alpha}}
\right\}\ . \nn
\eea
The upper (lower) sign corresponds to $-$ ($+$) sign in (\ref{EOS7c}) and
$- \frac{4(1-\alpha)}{1-2\alpha}C > 0$ $\left(- \frac{4(1-\alpha)}{1-2\alpha}C < 0\right)$
in (\ref{EOSa9a}),
$- \frac{4(1-\alpha)}{1-2\alpha}\tilde C > 0$
$\left(- \frac{4(1-\alpha)}{1-2\alpha}\tilde C < 0\right)$
in (\ref{EOSa9b}).

Note that the such scalar theory exactly reproduces the singularity and/or
behavior in the scale factor (\ref{EOS5}) and the Hubble rate (\ref{EOS6}), which is generated
by the EoS (\ref{EOS1}) of previous section dark fluid. This shows that
such a perfect fluid with the EOS (\ref{EOS1}) can
be realized by the scalar field with specific potential. The important
lesson of
this presentation is that big class of ideal fluid as well as scalar
field dark energies brings the future universe to the finite-time
singularity of one of the four types under consideration.

\section{Singularities of spherically-symmetric spacetime filled with dark
energy}

In this section we show that finite-time singularities of dark energy
models in above two sections manifest themselves as radius singularities
of
spherically-symmetric spacetime filled with such dark energies.
In \cite{Kobayashi:2008tq}, it has been pointed that
 curvature singularity is realized inside the relativistic star for
a
viable
class of $f(R)$-gravities (for review of viable, realistic models of that
sort, see \cite{Nojiri:2008nt}).
Since such spherically-symmetric solution with a naked singularity is
inconsistent, this result indicates that large star (or even planet) could
not be formed, or such a relativistic star could be unstable
in such a theory.

It is known that viable modified gravity also shows all four above types
of
finite-time singularity\cite{singularity}. Hence, it is natural to
expect that qualitatively similar situation should be typical for any dark
energy which brings the universe to finite-time singularity.
Motivated with these observations, we investigate the EoS dark fluid,
which generates curvature
singularity in spherically-symmetric spacetime. Especially in this
section, we investigate the singularity, which is generated
for a finite (non-vanishing) value of the radius. The singularity at the origin (vanishing
radius) is investigated in the next section.

Let us first consider what kind of (perfect) fluid could generate a
singularity.
We concentrate on the singularity which occurs for a finite radius for
spherically
symmetric solution.
Assume the metric has the following form:
\be
\label{sss1}
ds^2 = - \e^{\nu(r)} dt^2 + \e^{2\lambda(r)} dr^2 + r^2 d\Omega_{2}^2\ .
\ee
Here $d\Omega_{2}^2$ expresses the metric of two-dimensional sphere.
Eq.(\ref{sss1}) expresses the arbitrary spherically symmetric and static
space-time also in the presence of
some matter, that corresponds to the inside of the star or planet.
Then the Einstein equations have the following form:
\bea
\label{sss2}
\frac{1}{r}\frac{d\lambda}{dr} + \frac{\e^{2\lambda} - 1}{r^2} &=& \kappa^2 \rho \e^{2\lambda}\ ,\\
\label{sss3}
\frac{1}{r}\frac{d\nu}{dr} - \frac{\e^{2\lambda} - 1}{r^2} &=& \kappa^2 p_r \e^{2\lambda}\ ,\\
\label{sss4}
\frac{d^2 \nu}{dr^2} + \left(\frac{d\nu}{dr} - \frac{d\lambda}{dr}\right)
\left(\frac{d\nu}{dr} + \frac{1}{r}\right) &=& \kappa^2 p_a \e^{2\lambda}\ .
\eea
Here $\rho$ is the energy density and $p_r$ and $p_a$ are the radial and angular components of the
pressure.
In the following, as for the usual perfect fluid, we assume $p_r=p_a$. If
one does not impose this
assumption, we can consider general types of singularity.
By combining (\ref{sss3}) and (\ref{sss4}) and deleting $p=p_r=p_a$, it
follows
\be
\label{sss5}
0= - \frac{d^2 \nu}{dr^2} - \left(\frac{d\nu}{dr}\right)^2
+ \frac{d\nu}{dr} \left(\frac{1}{r} + \frac{d\lambda}{dr}\right)
+ \frac{1}{r}\frac{d\lambda}{dr} - \frac{\e^{2\lambda} - 1}{r^2}\ .
\ee
The following kind of singularities at $r=r_0$ may be now considered:
\be
\label{sss6}
\lambda(r) = \lambda_0 + \lambda_1 \ln \frac{ r - r_0}{r_0}
+ \sum_{n=2}^\infty \lambda_n \left( r - r_0 \right)^{n-1}\ ,\quad
\nu(r) = \nu_0 + \nu_1 \ln \frac{ r - r_0}{r_0}
+ \sum_{n=2}^\infty \nu_n \left( r - r_0 \right)^{n-1}\ .
\ee
By substituting (\ref{sss6}) into (\ref{sss5}), one obtains
\bea
\label{sss7}
&& \frac{\nu_1 - \nu_1^2 + \nu_1 \lambda_1}{\left(r - r_0\right)^2}
+ \frac{\frac{\nu_1 + \lambda_1}{r} - 2\nu_1 \nu_2 + \nu_1 \lambda_2 + \nu_2 \lambda_1}{r-r_0} \nn
&& - \nu_2^2 - \lambda_2 \nu_2 + \frac{\nu_2 + \lambda_2}{r} + \frac{1}{r^2}
+ {\cal O}\left(\left(r - r_0\right)\right)
= - \frac{\e^{2\lambda_0 + \sum_{n=2}^\infty \lambda_n \left( r - r_0 \right)^{n-1}}}{r^2}
\left(\frac{r- r_0}{r_0}\right)^{2\lambda_1}\ .
\eea
%In order that the equation (\ref{sss7}) has a solution, we find $2\lambda_1$ should be an integer.
Since the l.h.s. in (\ref{sss7}) contains only the power $\left(r - r_0\right)^m$,
where $m$ is an integer greater than or equal to $-2$, $2\lambda_1$ should be also an integer greater
than or equal to $-2$.
Furthermore if we assume $2\lambda_1 = -2$, from the coefficients of $\left(r - r_0\right)^{-2}$, we
find $0 = \nu_1^2 + \e^{2\lambda_0}$, which is inconsistent since the r.h.s. is positive definite.
Therefore $2\lambda_1$ must be an integer greater than or equal to $-1$; $2\lambda_1 \geq -1$.
Then from the coefficients of $\left(r - r_0\right)^{-2}$, again, we find
\be
\label{sss8}
0 = \nu_1 \left(1 - \nu_1 + \lambda_1\right)\ ,\quad \mbox{that is,} \quad
\nu_1 = 0\ \mbox{or}\ \nu_1 = \lambda_1 + 1\ .
\ee
If $\nu_1=\lambda_1=0$, there is no singularity and we do not consider this case.
In case $\nu_1=0$ and $2\lambda_1=-1$, from the coefficients of $\left(r - r_0\right)^{-2}$ in (\ref{sss7}),
we find
$-\frac{1}{2r_0} - \frac{\nu_2}{2} - \frac{\e^{2\lambda_0}}{r_0}=0$.
In case $\nu_1=0$ and $2\lambda_1\geq 1$, one finds
\be
\label{sss10}
\nu_2 = - \frac{1}{r_0}\ .
\ee
The case $\nu_1 \neq 0$ and $\nu_1 = \lambda_1 + 1 = 1/2$ corresponds to the black hole, where
$r_0$ corresponds to the horizon radius and it follows
$- \frac{3}{2}\nu_2 + \frac{\lambda_2}{2} - \frac{\e^{2\lambda_0}}{r_0}=0$.
In case $\nu_1=1$ and $\lambda_1=0$, we find
$\frac{1}{r_0} - 2\nu_2 + \lambda_2 = 0$.
In case $\nu_1 =\lambda_1 + 1 \geq 3/2$, one gets
$\frac{\nu_1 + \lambda_1}{r_0} - 2\nu_1 \nu_2 + \nu_1 \lambda_2 + \nu_2 \lambda_1 = 0$.

We now investigate how $\rho$ and $p$ behave.
In case $\nu_1=0$ and $2\lambda_1 = -1$, from Eqs. (\ref{sss2}) and (\ref{sss3}),
$\rho$ and $p$ are not singular and behave as
$\rho \sim - p \sim \frac{1}{\kappa^2 r_0^2}$.
Therefore there could not be the  singularity at $r=r_0$.
In case $\nu_1 = 0$ and $2\lambda_1= n \geq 1$, Eqs. (\ref{sss2}) and (\ref{sss3}) give
$\rho \sim \frac{2n \e^{-2\lambda_0}r_0^{n-1}}{\kappa^2 \left(r - r_0\right)^{n+1}}$,
$p \sim - \frac{\e^{-2\lambda_0}r_0^{n-2}}{\kappa^2 r_0 \left(r - r_0\right)^n}$.
Here relation (\ref{sss10}) is used. Then $\rho$ and $p$ satisfy the
following asymptotic
equation of state (EoS):
\be
\label{sss16}
p \sim - C \rho^{\frac{n}{n+1}}\ .
\ee
Here $C$ is a positive constant. Since $\rho$ and $p$ diverge at $r=r_0$, there is a curvature
singularity at $r=r_0$.
As we will see soon, the EoS (\ref{sss16}) corresponds to that in (\ref{EoSt3}),
which generates type III singularity.

In case $\lambda_1 = 0$ and $\nu_1=1$ case, we find
$\rho \sim \frac{1}{\kappa^2}\left(\frac{2\e^{-2\lambda_0} \lambda_2}{r_0} + \frac{1}{r_0^2}
 - \frac{\e^{-2\lambda_0}}{r_0^2} \right)$,
$p \sim \frac{2\e^{-2\lambda_0}}{\kappa^2 r_0 \left(r - r_0\right)}$.
Here $\rho$ is finite although $p$ diverges at $r=r_0$. The divergence of $p$ generates
the curvature singularity. The EoS has the following form:
$p \left(\rho - \rho_0\right) \sim \mbox{const.}$.

In case $2\lambda_1=n\geq 0$ and $\nu_1 = \lambda_1 + 1$,
we find
\be
\label{sss18}
\rho \sim \frac{n\e^{-2\lambda_0}r_0^{n-1}}{\kappa^2 \left(r - r_0\right)^{n+1}}\ ,\quad
p \sim \frac{(n+2) \e^{-2\lambda_0}r_0^{n-1}}{\kappa^2 \left(r - r_0\right)^{n+1}}\ .
\ee
and therefore the following asymptotic EoS:
\be
\label{sss19}
p= \left( 1 + \frac{2}{n}\right) \rho\ .
\ee
The divergence of $\rho$ and $p$ at $r=r_0$ means the curvature singularity.
Since the EoS parameter $w\equiv p/\rho = 1 + 2/n$ is positive, the EoS
does not
generate any finite-time singularity and corresponds to $\alpha=1$ and $-1
+ A = 1 + 2/n$.

Hence, rather general case with the singularity occurence for finite
$r$ is investigated.
We found here essentially two types of singularity expressed by the
EoS fluid
(\ref{sss16}) or (\ref{sss19}).
These EoS fluids could be compared with the asymptotic EoS dark fluid
generating
the finite time singularity
(\ref{EoSt1}), (\ref{EoSt2}), and (\ref{EoSt3}).
First, one notices that EoS fluid (\ref{sss19}) for finite
radius singularity
does not exist for dark fluid generating the finite time singularity as in
(\ref{EoSt1}), (\ref{EoSt2}), and (\ref{EoSt3}).
(However, it could be that other dark fluid generating future singularity
shows up radius singularity in this example).
Eq.(\ref{EoSt1}) has a similar structure but in (\ref{EoSt1}), $w\equiv p/\rho \leq -1$ although $w>1$
in (\ref{sss19}).
The EoS (\ref{sss16}) has a similar structure with those in (\ref{EoSt1}) and (\ref{EoSt1}) if we
identify
\be
\label{sss19b}
\alpha = \frac{n}{n+1}\ .
\ee
In case of (\ref{EoSt1}), $\alpha$ is negative but in case of (\ref{EoSt3}), $\alpha$ is positive.
Then the fluid with EoS (\ref{EoSt3}), which generates type III
singularity, may generate finite radius
singularity. Note, however, we do not have the explicit proof that the fluid
generating finite-time singularity always
generate finite radius singularity, and vice versa. That depends also from
the distance between energy scale which is typical for future singularity
and the star/black hole formation energy scale.
 The important lesson of our study is that dark fluid generating
finite-time singularity may manifest it also as radius singularity of
spherically-symmetric solution (i.e. stars, black holes, etc).
%%%%%%%%%

In the above analysis, it is assumed that $\lambda(r)$ and $\nu(r)$ are
real and
therefore $\e^{2\lambda(r)}$ and $\e^{2\nu(r)}$ are positive. As we have only investigated
the local behavior, the global structure of the spacetime is not so clear.
Since
$\e^{2\lambda(r)}$ and $\e^{2\nu(r)}$ are positive, however, the number of the horizons must be
even and in principle, the singularity can be naked. If there is a usual matter, whose distribution
is also spherically symmetric, as a main component of the star, the singularity might be hidden
inside the matter. If the radius of the matter becomes smaller than the radius of the singularity,
however, due to the strong gravitational force near the singularity, the
matter can shrink very rapidly,
which may produce the strong radiation, say, X-ray.

If $\e^{2\lambda(r)}$ and/or $\e^{2\nu(r)}$ are negative near the singularity, however, there must be
horizon, where $\e^{2\lambda(r)}$ and/or $\e^{2\nu(r)}$ vanish, since $\e^{2\lambda(r)}$ and
$\e^{2\nu(r)}$ are positive in a region far from the star. We should note that if $\e^{2\lambda_0}$ is
negative, $0 = \nu_1^2 + \e^{2\lambda_0}$ has non-trivial solution even if $2\lambda_1 = -2$.
In case of (\ref{sss18}), where $2\lambda_1=n\geq 0$ and $\nu_1 = \lambda_1 + 1$, if
$\e^{2\lambda_0}$ is negative, $\rho$ becomes negative and inconsistent. Therefore the singularity
with $2\lambda_1=n\geq 0$ and $\nu_1 = \lambda_1 + 1$ could be forbidden
in some cases (compare with singularities classification in
ref.\cite{singularityBH}).

\section{Central singularity of spherically-symmetric spacetime filled by
dark energy}

In the previous section, we have investigated the singularity,
which occurs at the finite
(non-vanishing) radius.
In this section, we consider the singularity at the center of the
spherically-symmetric
solution and clarify what kind of EoS dark fluid could generate such a
singularity.
It is also shown that there are some exact solutions.
Since such a solution with a naked singularity is inconsistent, even if
the singularity exists
at the center of the star, this result indicates that large star/planet
could not be formed, or such a relativistic star is unstable.

Assume that the energy density $\rho$ and the pressure of the perfect
fluid
only depend on the radial coordinate $r$, then the conservation law gives
\be
\label{sp2}
0 = \frac{d p}{d r} + \frac{d\nu}{d r}\left(p+\rho\right)=0\ ,
\ee
which could be derived by using (\ref{sss2}), (\ref{sss3}), and (\ref{sss4}).
Especially for the perfect fluid with a constant EoS parameter $w$:
$p=w\rho$, we find $\rho = \rho_0 \e^{- \frac{1+w}{w}\nu(r)}$.
When $\nu$ is a monotonically increasing function of the radial coordinate $r$, as
in the Schwarzschild metric, $\rho$ becomes a monotonically decreasing function
of $r$ if $w>0$ or $w<0$ but $\rho$ becomes a monotonically increasing function
if $0>w>-1$. Then the gravity effectively acts as a repulsive force when $0>w>-1$.

For later convenience, it is convenient to write the EoS in the
following form:
\be
\label{sp4}
\rho + p = F(p)\ .
\ee
Here $F(p)$ is an appropriate function. Then (\ref{sp2}) gives
\be
\label{sp5}
\nu = \int \frac{dp}{F(p)}\ .
\ee
We should note that (\ref{sss3}) with $p_r=p=p_a$ can be rewritten as
\be
\label{sp9}
\lambda = \frac{1}{2} \ln \left\{\left(\frac{2}{r}\frac{d\nu}{d r} + \frac{1}{r^2} \right)
\left(\kappa^2 p + \frac{1}{r^2} \right)^{-1} \right\}\ .
\ee
Then by using (\ref{sss2}) and (\ref{sp9}), one can delete $\lambda$ and
obtain
\be
\label{sp10}
0 = \frac{\frac{2}{r}\frac{d^2\nu}{d r^2} - \frac{2}{r^2}\frac{d\nu}{d r} - \frac{2}{r^3}}
{\frac{2}{r}\frac{d\nu}{d r} + \frac{1}{r^2}}
 - \frac{\kappa^2 \frac{d p}{d r} - \frac{2}{r^3}}{\kappa^2 p + \frac{1}{r^2}}
 - \frac{\left(\frac{2}{r}\frac{d\nu}{d r} + \frac{1}{r^2} \right)
\left( \kappa^2 \rho - \frac{1}{r^2} \right) }{\kappa^2 p + \frac{1}{r^2}}
 - \frac{1}{r}\ .
\ee
By using (\ref{sp4}) and (\ref{sp5}), we may delete $\nu$ and $\rho$ in (\ref{sp10})
\be
\label{sp11}
0 = \frac{ - \frac{2}{r F(p)} \frac{d^2 p}{d r^2}
+ \frac{2}{r F(p)^2} \frac{d F(p)}{d p} \left(\frac{d p}{d r}\right)^2
+ \frac{2}{r^2 F(p)}\frac{d p}{d r} - \frac{2}{r^3}}
{ - \frac{2}{r F(p)}\frac{d p}{d r} + \frac{1}{r^2}}
 - \frac{\kappa^2 \frac{d p}{d r} - \frac{2}{r^3}}{\kappa^2 p + \frac{1}{r^2}}
 - \frac{2}{F(p)}\frac{d p}{d r}
 - \frac{ - 2 \kappa^2 \frac{d p}{d r} + \frac{\kappa^2 F(p)}{r}}
{\kappa^2 p + \frac{1}{r^2}}\ .
\ee
Eq.(\ref{sp11}) can be regarded as the differential equation for $p=p(r)$
for a given $F(p)$.
With known solution of (\ref{sp11}) with respect to $p=p(r)$, the
explicit form of
$\nu$ may be found by using (\ref{sp5}). Then by using (\ref{sp9}), we
obtain $\lambda$. Finally Eq.(\ref{sss4})
might determine some of the integration constants.

As an example, we consider the perfect fluid with a constant equation of state (EoS)
parameter $w$. It follows
$F(p) = \alpha p$, $\alpha \equiv 1 + \frac{1}{w}$.
Then one of the exact solutions in (\ref{sp11}) could be given by
\be
\label{sp13}
p = \frac{p_0}{r^2}\ ,\quad p_0 = \frac{4}{\kappa^2\left(\alpha^2 + 4 \alpha - 4\right)}
 = \frac{4w^2}{\kappa^2\left(w^2 + 6 w + 1\right)}\ ,
\ee
which gives
\be
\label{sp14}
\nu = \frac{2}{\alpha}\ln \frac{r}{r_0} = \frac{2w}{w + 1} \ln \frac{r}{r_0}\ ,\quad
\lambda = \lambda_0 \equiv \frac{1}{2} \ln \left( \frac{\alpha^2 + 4\alpha - 4}{\alpha^2}\right)
= \frac{1}{2} \ln \left( \frac{w^2 + 6w + 1}{\left(w+1\right)^2} \right)\ .
\ee
When $r\sim 0$, the scalar curvature $R$ behaves as
$R = - \e^{-2\lambda_0} \left(\frac{8}{\alpha^2} + \frac{4}{\alpha}\right) \frac{1}{r^2}$.
Hence, except $\alpha=-2$, that is, $w = - 1/3$, there is surely (naked)
singularity at $r=0$.

We now consider more general case, as in (\ref{sss6}),
\be
\label{ssss1}
\lambda(r) = \lambda_0 + \lambda_1 \ln \frac{r}{r_0}
+ \sum_{n=2}^\infty \lambda_n r^{n-1}\ ,\quad
\nu(r) = \nu_0 + \nu_1 \ln \frac{r}{r_0}
+ \sum_{n=2}^\infty \nu_n r^{n-1}\ .
\ee
By substituting (\ref{ssss1}) into (\ref{sss5}), one gets
\bea
\label{ssss2}
&& \frac{\nu_1 - \nu_1^2 + \nu_1 + \nu_1 \lambda_1 + \lambda_1 + 1}{r^2}
+ {\cal O}\left(r^{-1}\right)
= - \frac{\e^{2\lambda_0 + \sum_{n=2}^\infty \lambda_n \left( r - r_0 \right)^{n-1}}}{r^2}
\left(\frac{r}{r_0}\right)^{2\lambda_1}\ .
\eea
Since the l.h.s. is only given by the integer power, in order that there could be a solution,
$2\lambda_1$ should be an integer $n$: $2\lambda_1 = n$.
We also find, if $n<0$, there is no solution in (\ref{ssss2}), therefore $n\geq 0$.

When $n=0$, by requiring the term with $r^{-2}$ to vanish, we obtain
$\e^{2\lambda_0}=\left(\sqrt{2} + 1 - \nu_1\right)\left(\nu_1 + \sqrt{2} - 1\right)$.
Since the r.h.s. is positive, $- \sqrt{2} + 1 < \nu_1 < \sqrt{2} + 1$.
When $n=0$ case, (\ref{sss2}) and (\ref{sss3}) gives
\be
\label{ssss6}
\kappa^2 \rho \sim \frac{\nu_1^2 - 2\nu_1}{\left(\nu_1^2 - 2\nu_1 + 1\right)r^2}\ ,\quad
\kappa^2 p \sim - \frac{\nu_1^2 + 3 \nu_1}{\left(\nu_1^2 - 2\nu_1 + 1\right)r^2}\ ,
\ee
which gives the following EoS parameter,
\be
\label{ssss7}
w = - \frac{\nu_1 + 3}{\nu_1 - 2}\ .
\ee
Unless $\nu_1=0$, $\rho$ and/or $p$ diverge at $r=0$ and therefore there is a curvature singularity
there. As we require $\rho$ could be non-negative, we find a constraint $\nu_1<0$ or $\nu_1>2$.
Combining this constraint with the previous one $- \sqrt{2} + 1 < \nu_1 < \sqrt{2} + 1$,
we find
\be
\label{ssss8}
- \sqrt{2} + 1 < \nu_1 < 0\ \mbox{or}\ 2< \nu_1 < \sqrt{2} + 1\ .
\ee
As $w$ in (\ref{ssss7}) is constant, this case corresponds to $\alpha=1$ case in (\ref{EOS1}).
Combining (\ref{ssss7}) with (\ref{ssss8}), one gets
$\frac{3}{2}<w<2+3\sqrt{2}$, $-3\sqrt{2} + 2 > w > -\infty$.
In the former case, $w$ is positive and therefore, there does not occur
any finite-time singularity.
In the latter case, however, $w$ is negative and $-3\sqrt{2} + 2 < -1$ and therefore this case
corresponds to the Big Rip or Type I singularity.

On the other hand, when $n\geq 1$, we find
\be
\label{ssss5}
%\lambda_1 = \frac{\nu_1^2 - 2\nu_1 - 1}{\nu_1 + 1}\ .
\nu_1 = \frac{4 - n \pm \sqrt{ n^2 + 32}}{4}
\ee
Then from (\ref{sss2}) and (\ref{sss3}), we obtain
\be
\label{ssss9}
\kappa^2 \rho \sim \e^{-n}\left(\frac{r}{r_1}\right)^{-n}\frac{\lambda_1 - 1}{r^2}
= \e^{-n}\left(\frac{r}{r_1}\right)^{-n}\frac{n - 2}{2 r^2}
\ ,\quad
\kappa^2 p \sim \e^{-n}\left(\frac{r}{r_1}\right)^{-n}\frac{\nu_1 - 1}{r^2}
= \e^{-n}\left(\frac{r}{r_1}\right)^{-n}\frac{- n \pm \sqrt{n^2 + 32}}{4r^2}\ ,
\ee
which gives the following EoS parameter
\be
\label{ssss10}
w= \frac{\nu_1 - 1}{\lambda_1 - 1} = \frac{- n \pm \sqrt{n^2 + 32}}{2(n-2)}\ ,
\ee
which is constant again and corresponds to $\alpha=1$ case in (\ref{EOS1}).
Here Eq.(\ref{ssss5}) is used.
In order that $\rho$ is positive, one finds $n\geq 2$. When $n=2$, $\rho$
vanishes and
\be
\label{ssss11}
\rho = 0\ ,\quad \kappa^2 p \sim - \frac{2\e^{-2}\left(\frac{r}{r_1}\right)^{-n}}{r^2}\, ,\
\frac{\e^{-2}\left(\frac{r}{r_1}\right)^{-n}}{r^2}\ .
\ee
Since $\rho$ is finite although $p$ diverges, this case corresponds to Type II
singularity, which corresponds to $\alpha<0$ in the EoS (\ref{EOS1}).
In the upper sign ($+$-sign) in (\ref{ssss10}), $w$ is positive and therefore
this case does not correspond to any finite-time singularity.
In the lower sign ($-$-sign) in (\ref{ssss10}), however, $w$ is negative
and it corresponds to
Type I singularity.

Since $\rho$ and/or $p$ diverge at $r=0$ in above cases, there is a
curvature singularity there.
If usual matter is a main component of the star, the singularity could be hidden even if
the singularity could be naked when there is no matter.
When the usual matter is stripped, if the interaction between the usual matter and the anomalous
matter, which could generate the singularity, is weak, the naked singularity could appear.
Due to the strong gravitational force, the singularity could attract the matter around the singularity
and strong radiation cold be generated.

We have investigated rather general singularities. In order to treat further general cases,
one can now consider a reconstruction program. When we require $p$ behaves
as
\be
\label{sp16}
p=P(r)\ ,
\ee
one can try to find $F(p)$ which realizes (\ref{sp16}).
Defining a function $G(p)$ by
\be
\label{sp17}
\frac{d G(p)}{dp} = \frac{1}{F(p)}\ ,
\ee
by using (\ref{sp16}), Eq.(\ref{sp11}) can be rewritten as
\be
\label{sp18}
0 = \frac{dG}{dr}\frac{d^2 G}{dr^2}
 - \frac{\kappa^2 r^3 P'(r) + 2}{r\left(\kappa^2 r^2 P(r) + 1\right)}\left(\frac{dG}{dr}\right)^2
 - 2 \left(\frac{dG}{dr}\right)^3 %\nn
+ \left\{ \frac{1}{r^2}
 - \frac{3\kappa^2 r^3 P'(r) - 2}{2r^2 \left(\kappa^2 r^2 P(r) + 1\right)} \right\}
\frac{d G}{d r} + \frac{\kappa^2 P'(r)}{2\left(\kappa^2 r^2 P(r) + 1\right)}\ .
\ee
Eq.(\ref{sp18}) is the first order differential equation for $dG/dr$. By solving (\ref{sp18}),
one may find $G$ as a function of $r$: $G = \tilde G (r)$. Then by using
(\ref{sp16}), we
get $G$ as a function of $p$ by $G(p) = \tilde G \left( P^{-1}(p) \right)$.
Here $P^{-1}(p)$ is an inverse function of $p=P(r)$. Using (\ref{sp4}) and
(\ref{sp17}), we
find that the EoS is given by
\be
\label{sp20}
1 = \left(\rho + p\right) \int^p dp'
\tilde G \left( P^{-1}(p') \right)\ .
\ee
Using the above equation, one may find the EOS dark fluid generating
general singularity expressed by (\ref{sp16}).

\section{Discussion}

In summary, we investigated dark fluid/scalar DE universe where all four
known types of future singularities occur for the corresponding choice of
EoS parameters/scalar potentials. The energy density/pressure of such DE
is singular
in the spherical spacetime at finite radius or at the center.
As a result, the relativistic stars and/or  black holes in
such universe may become unstable which makes the corresponding DE model to
be problematic as realistic theory for current universe acceleration.
One way to overcome this problem is to modify the initial EoS in such a
way that finite-time singularity does not occur or occurs in far remote
future so that it does not influence star/BH formation processes.

The contributions to the FRW equations
from usual matter could be neglected near  the singularity. Then
combining the FRW equations with (\ref{sp4}), we find $-2\dot H/\kappa^2 \sim F(p)$.
Hence, if the absolute value of $F(p)$ is bounded, there does not occur
Type I, II, or III
singularities where $\dot H$ diverges. One can achieve this by extra
modification of the corresponding EoS. From another side, in the same way
as in second work of ref.\cite{singularity} one can add inhomogeneous
(gravitational) terms which are relevant at the early universe,
 to the EoS so that the future universe becomes regular. However, the
introduction of such terms effectively means the modification of gravity
(for general review of various modified gravity models see \cite{review}).
Eventually, this brings us back to modified gravity as consistent DE
model.

\section*{Acknowledgments}

The work by S.N. is supported in part by the Ministry of Education,
Science, Sports and Culture of Japan under grant no.18549001 and Global
COE Program of Nagoya University provided by the Japan Society
for the Promotion of Science (G07).
The work by S.D.O. is supported in part by MICINN (Spain) projects
FIS2006-02842 and PIE2007-50I023 and by LRSS project N.2553.2008.2.

\end{document}